\documentclass[doublespacing]{elsart}

\usepackage{graphics}
\usepackage{graphicx}
\usepackage{epsfig}
\usepackage{color}
\usepackage{subfigure}

\usepackage{amssymb}
\usepackage{amsmath}
\usepackage{amsfonts}
\usepackage{stmaryrd}
\usepackage{txfonts}
\usepackage{mathrsfs}
\usepackage{latexsym, bm}
\newtheorem{theorem}{Theorem}

\newtheorem{definition}{Definition}
\newtheorem{lemma}{Lemma}
\newtheorem{corollary}{Corollary}

\newtheorem{remark}{Remark}
\newtheorem{example}{Example}

\begin{document}


\begin{frontmatter}


\title{Consensus of switched multi-agent systems\thanksref{label1}}
\thanks[label1]{This work was supported by 973 Program (Grant No. 2012CB821203), NSFC (Grant
Nos. 61020106005, 61375120 and 61304160) and the Fundamental Research Funds for
the Central Universities (Grant Nos. JB140406 and NSIY211416).
}

\author[xidian]{Yuanshi Zheng }, \ead{zhengyuanshi2005@163.com}
\author[xidian]{Jingying Ma}, \ead{majy1980@126.com}
\author[beida]{Long Wang \corauthref{cor1}}  \ead{longwang@pku.edu.cn}

\corauth[cor1]{Corresponding author : Long Wang }
\address[xidian]{ Center for Complex Systems, School of Mechano-electronic Engineering, Xidian University, Xi'an 710071, China}
\address[beida]{Center for Systems and Control, College of Engineering, Peking University, Beijing 100871, China}

\begin{abstract}
In this paper, we consider the consensus problem of switched multi-agent system composed of continuous-time and discrete-time subsystems. By combining the classical consensus protocols of continuous-time and discrete-time multi-agent systems, we propose a linear consensus protocol for switched multi-agent system. Based on the graph theory and Lyapunov theory, we prove that the consensus of switched multi-agent system is solvable under arbitrary switching with undirected connected graph, directed graph and switching topologies, respectively. Simulation examples are also provided to demonstrate the effectiveness of the theoretical results.
\end{abstract}

\end{frontmatter}

\newpage
\section{\bf Introduction}\label{s-introduction}

In the past decade, multi-agent coordination has made great progress due to the rapid developments of computer science and communication technologies. It has received a major attention of multidisciplinary researchers including system control theory, mathematics, biology, statistical physics and so on. This is partly due to its broad applications in many fields, such as formation control, flocking, synchronization and target tracking of robots, social insects, complex networks, sensor networks, etc \cite{Saber06,chu06,xiao09,ji09}.

Consensus problem is an important and challenging research topic in multi-agent coordination, which is to design appropriate control input based on local information that enables all agents to reach an agreement on consistent quantity of interest. Vicsek et al. \cite{vicsek95} proposed a simple model for a group of self-driven particles and demonstrated by simulation that the system will synchronize if the population density is large. By virtue of graph theory, Jadbabaie et al. \cite{jabdabaie03} explained the consensus behaviour of Vicsek model theoretically and shown that the consensus can be achieved if the union of interaction graph are connected frequently enough. Olfati-Saber and Murray \cite{sabertac04} discussed the consensus problem of multi-agent systems with switching topologies and time-delays in a continuous-time (CT) model and obtained some useful results for solving the average consensus problem. Ren and Beard \cite{ren05} extended the results given in \cite{sabertac04} and presented some more relaxable conditions for consensus with switching topologies. With the development of this issue, lots of new results were given out with different models and consensus protocols. Hong et al. \cite{hong06} considered the multi-agent consensus with an active leader and variable topology. By utilizing the pre-leader-follower decomposition, Wang and Xiao \cite{xiao06-1} studied the state consensus of discrete-time (DT) multi-agent systems with switching topologies and bounded time-delays. Based on linear matrix inequality (LMI) approach, Sun et al. \cite{Sun08} studied the average consensus of multi-agent systems with switching topologies and time-varying delays. Lin and Jia \cite{lin09} considered the consensus of DT second-order  multi-agent systems with switching topologies and nonuniform time-delays. \cite{ni2010} investigated the leader-following consensus of high-order multi-agent systems with fixed and switching topologies. Zheng and Wang proposed a heterogeneous multi-agent systems which is composed of first-order and second-order integrator agents \cite{zheng11-2} and studied the consensus problem under directed fixed and switching topologies \cite{zheng12-3}. Other research topics for consensus with switching topologies were considered, such as asynchronous consensus \cite{Xiao08-1}, finite-time consensus \cite{jiang09}, stochastic consensus \cite{li10}, group consensus \cite{yu10}, sampled-date based consensus \cite{gao11} and so on. To date, CT/DT multi-agent consensus has been wildly analyzed with time-varying topologies by using graph theory, Lyapunov theory, LMI approach, etc. For more details, one can refer to survey papers \cite{Olfati-Saber07} and the references therein.

It should be noted that all the aforementioned references were concerned with multi-agent consensus under switching topologies, i.e. the multi-agent system is composed of only CT subsystems or only DT subsystems. However, it is easy to find many applications of switched multi-agent system which is composed of both CT and DT subsystems. For example, in a CT switched multi-agent systems, if we sometimes use computer to activate all the agents in a DT manner, then the switched multi-agent system is composed of both CT and DT subsystems. In \cite{zhai04}, Zhai et al. studied the stability of switched systems which are composed of a DT subsystem a DT subsystem. Some algebraic conditions are given for solving the stability problem under arbitrary switching. Inspired by the stability analysis for switched system in \cite{zhai04}, we try to investigate the consensus problem of switched multi-agent system composed of continuous-time and discrete-time subsystems. By combining the classical consensus protocols of CT and DT multi-agent systems, we propose a linear consensus protocol for switched multi-agent system. The main aim of this paper is to obtain the graphic criterions for consensus of switched multi-agent system in different networks. Firstly, by utilizing of graph theory and Lyapunov theory, we obtain that the consensus can be achieved with arbitrary switching under undirected connected graph if the sampling period $0<h<\frac{2}{\lambda_n}$. Secondly, the consensus under directed graph is also analyzed by using the previous results of CT and DT multi-agent consensus. Thirdly, we give a sufficient condition for the consensus of switched multi-agent system with switching topologies.

The rest of this paper is organized as follows. In Section \ref{s-preliminaries}, we present some notions in graph theory and propose the switched multi-agent system. In Section \ref{s-Main results}, we give the main results. In Section \ref{s-Simulations}, numerical simulations are given to illustrate the effectiveness of theoretical results. Finally, some conclusions are drawn in Section \ref{s-Conclusion}.

The following notations will be used throughout this paper: $\mathbb{Z}$ and $\mathbb{R}$ denote the set of integer and real  number, $\mathbb{R}^N$ denotes the $N-$dimensional real vector space. $\mathcal{I}_{n}=\{1,2, \dots,\ n\}$. For a given vector or matrix $A,$ $A^{T}$ denotes its transpose.
$\mathbf{1}_{n}$  is a vector with elements being all ones. $I_{n}$ is the $n\times n$ identity matrix.
$0$ $(0_{m\times n})$ denotes an all-zero vector or matrix with compatible dimension (dimension $m\times n$). Given a complex number $\lambda\in \mathbb{C}$, $Re(\lambda)$, $Im(\lambda)$ and $|\lambda|$ are the real part, the imaginary part and the modulus of $\lambda$, respectively.


\section{\bf Preliminaries}\label{s-preliminaries}

\subsection{Graph theory}\label{s-graph theory}

In this subsection, we first introduce some basic concepts and results about graph theory. For more details, please refer to \cite{Godsil01}.

A weighted directed graph $\mathscr{G}(\mathscr{A})=(\mathscr{V},\mathscr{E},\mathscr{A})$ of order $n$
consists of a vertex set $\mathscr{V}=\{s_{1}, s_{2}, \cdots, s_{n}\}$,  an edge set
$\mathscr{E}=\{e_{ij}=(s_{i}, s_{j})\}\subset \mathscr{V}\times \mathscr{V}$ and a nonnegative
matrix $\mathscr{A}=[a_{ij}]_{n\times n}$.  A directed path between two distinct vertices $s_{i}$ and $s_{j}$ is
a finite ordered sequence of distinct edges of $\mathscr{G}$ with the form $(s_{i}, s_{k_{1}}), (s_{k_{1}}, s_{k_{2}}), \cdots, (s_{k_{l}}, s_{j})$.
A directed tree is a directed graph, where there exists a vertex called the root such that there exists a unique directed path from this vertex to every other vertex. A directed spanning tree is a directed tree, which consists of all the nodes and some edges in $\mathscr{G}$.
If a directed graph has the property that $(s_{i}, s_{j})\in \mathscr{E} \Leftrightarrow (s_{j}, s_{i})\in \mathscr{E}$, the directed graph is called undirected. An undirected graph is said to be connected if there exists a path between any two distinct vertices of the graph. The degree matrix
$\mathscr{D}=[d_{ij}]_{n\times n}$ is a diagonal matrix with $d_{ii}=\sum_{j:s_j\in \mathscr{N}_{i}} a_{ij}$ and
the Laplacian matrix of the graph is defined as $\mathscr{L}=[l_{ij}]_{n \times n}=\mathscr{D}-\mathscr{A}.$ It is easy to see that $\mathscr{L}\mathbf{1}_{n}=0$. Thus, the eigenvalues of $\mathscr{L}$ can be denoted as $0=\lambda_1\leq\lambda_2\leq\cdots\leq\lambda_n$.
When $\mathscr{G}$ is a connected undirected graph, $\mathscr{L}$ is positive semi-definite and has a simple zero eigenvalue,  $\xi^{T}\mathscr{L}\xi=\frac{1}{2}\sum_{i,j=1}^{n}a_{ij}(\xi_{j}-\xi_{i})^2$ and $\min_{\xi\neq 0, \mathbf{1}_{n}^{T}\xi=0}\frac{\xi^{T}\mathscr{L}\xi}{\xi^{T}\xi}=\lambda_2 $ for any $\xi=[\xi_{1}, \xi_{2}, \cdots \xi_{n}]^{T}\in \mathbb{R}^{n}$.

\subsection{System model}\label{s-System model}

In this subsection, we propose the switched multi-agent system which is composed of a CT subsystem
\begin{equation}\label{m-CT}
   \dot{x}_{i}(t)=u_{i}(t), ~~~~i\in\mathcal{I}_{n},
  \end{equation}
and a DT subsystem
\begin{equation}\label{m-DT}
   x_{i}(t+1)=x_{i}(t)+u_{i}(t),  ~~~~i\in\mathcal{I}_{n},
  \end{equation}
where $x_{i}\in \mathbb{R}$ and $u_{i}\in \mathbb{R}$ are the position and control input of agent $i$, respectively. The initial condition is $x_{i}(0)=x_{i0}$. Let $x(0)=[x_{10}, x_{20}, \ldots, x_{n0}]^{T}$.

\begin{definition}\label{def-consensus}
The switched multi-agent system (\ref{m-CT}--\ref{m-DT}) is said to reach consensus if for any initial state, there exists $x^{*}$ (dependent on the initial state) such that
\[
\begin{aligned}
   \lim_{t \to \infty }\|x_{i}(t)-x^{*}\|=0, &&for~~~  i\in \mathcal{I}_{n}.  \\
\end{aligned}
\]
\end{definition}

The consensus protocols have been widely applied for the CT multi-agent system (\ref{m-CT}) and DT multi-agent system (\ref{m-DT}). We present the linear consensus protocol for the switched multi-agent system (\ref{m-CT}--\ref{m-DT}) as follows
\begin{equation}\label{linear consensus protocol}
u_{i}(t)=
\left\{
   \begin{aligned}
   &\sum_{j=1}^{n}a_{ij}(t)(x_{j}(t)-x_{i}(t)) &&for~~ CT~~ subsystem,\\
   &h\sum_{j=1}^{n}a_{ij}(t)(x_{j}(t)-x_{i}(t)), &&for~~ DT~~ subsystem,  \\
   \end{aligned}
   \right.
  \end{equation}
where $\mathscr{A}=[a_{ij}(t)]_{n\times n}$ is the weighted adjacency matrix associated with the graph $\mathscr{G}(t)$ at time instant $t$, $h>0$ is the sampling period.

Thus, the switched multi-agent system (\ref{m-CT}--\ref{m-DT}) with protocol (\ref{linear consensus protocol}) can be written as
\begin{subequations}\label{SMAS}
   \begin{align}
   &\dot{x}(t)=-\mathscr{L}(t)x(t), \label{SMAS-CT}\\
   &x(t+1)=(I_{n}-h\mathscr{L}(t))x(t).  \label{SMAS-DT}
   \end{align}
  \end{subequations}


\section{\bf Main results}\label{s-Main results}

In this section, the consensus problem of switched multi-agent system (\ref{SMAS}) will be considered  for network with fixed undirected graph, fixed directed graph and switching topologies, respectively.

Firstly, we consider the consensus of switched multi-agent system (\ref{SMAS}) in undirected graph with fixed topology, i.e., $\mathscr{L}(t)=\mathscr{L}$ and $\mathscr{L}^{T}=\mathscr{L}$ for any time $t$.

\begin{theorem}\label{T-undirected-fixed}
Suppose the communication network $\mathscr{G}$ is undirected and connected. Then, the switched multi-agent system (\ref{SMAS}) can solve the consensus problem under arbitrary switching if the sampling period $0<h<\frac{2}{\lambda_n}$.
\end{theorem}
{\bf Proof.} Let $c(t)=\frac{1}{n}\sum_{i=1}^{n}x_{i}(t)$. Since $a_{ij}=a_{ji}$ for all $i, j \in\mathcal{I}_{n}$, we have
\begin{equation}\label{dc(t)}
   \begin{aligned}
   \frac{dc(t)}{dt}=\frac{1}{n}\sum_{i=1}^{n}\frac{dx_{i}(t)}{dt}=-\frac{\mathbf{1}_{n}\mathscr{L}x(t)}{n}=0,
   \end{aligned}
  \end{equation}
and
\begin{equation}\label{c(t+1)}
   \begin{aligned}
   c(t+1)=\frac{1}{n}\sum_{i=1}^{n}x_{i}(t+1)=\frac{\mathbf{1}_{n}(I_{n}-h\mathscr{L})x(t)}{n}=\frac{1}{n}\sum_{i=1}^{n}x_{i}(t)=c(t).
   \end{aligned}
  \end{equation}
Therefore, $c(t)$ is time-invariant, i.e. $c(t)=c(0)$. Let $\delta(t)=x(t)-\mathbf{1}_{n}c(t)$, we have $\mathbf{1}^{T}_{n}\delta(t)=0$ and
\begin{subequations}\label{SMAS-delta}
   \begin{align}
   &\dot{\delta}(t)=-\mathscr{L}\delta(t), \label{SMAS-CT-delta}\\
   &\delta(t+1)=(I_{n}-h\mathscr{L})\delta(t).  \label{SMAS-DT-delta}
   \end{align}
\end{subequations}

We consider the common Lyapunov function $V(\delta(t))=\delta^{T}(t)\delta(t)$ for CT subsystem (\ref{SMAS-CT-delta}) and DT subsystem (\ref{SMAS-DT-delta}). Owing to $\min_{\xi\neq 0, \mathbf{1}_{n}^{T}\xi=0}\frac{\xi^{T}\mathscr{L}\xi}{\xi^{T}\xi}=\lambda_2$, in the period where CT subsystem (\ref{SMAS-CT-delta}) is activated, we have
\[
\begin{aligned}
 \dot{ V}(\delta(t))=-2\delta(t)^{T}\mathscr{L}\delta(t)\leq-2\lambda_2\delta(t)^{T}\delta(t)=-2\lambda_2V(\delta(t)),
\end{aligned}
\]
and in the period where DT subsystem (\ref{SMAS-DT-delta}) is activated, we obtain
\[
\begin{aligned}
 V(\delta(t+1))-V(\delta(t))&=\delta^{T}(t)(I_{n}-h\mathscr{L})^{2}\delta(t)-\delta^{T}(t)\delta(t)\\
 &=\delta^{T}(t)(-2h\mathscr{L}+h^{2}\mathscr{L}^{2})\delta(t)\\
 &\leq (-2h\lambda_2+h^{2}\lambda_2^{2})V(\delta(t)).
\end{aligned}
\]
Due to $0<h<\frac{2}{\lambda_n}$, we have $-2h\lambda_2+h^{2}\lambda_2^{2}<0$.

For any time $t>0$, we have $t=t_{c}+t_{d}$, where $t_{c}\in\mathbb{R}$ is the total duration time on CT subsystem (\ref{SMAS-CT-delta}) and $t_{d}\in\mathbb{Z}$ is the total duration time on DT subsystem (\ref{SMAS-DT-delta}). Let $k=1-2h\lambda_2+h^{2}\lambda_2^{2}$. Thus, we have $0<k<1$ and
\[
\begin{aligned}
 V(\delta(t))&\leq e^{-2\lambda_2t_{c}}k^{t_{d}}V(\delta(0))=e^{-2\lambda_2t_{c}}e^{-t_{d}ln({\frac{1}{k}})}V(\delta(0))\\
 &\leq e^{-2\alpha t}V(\delta(0)),
\end{aligned}
\]
where $\alpha=\min\{\lambda_2, \frac{ln(1/k)}{2}\}$, which implies $|\delta(t)|\leq e^{-\alpha t}|\delta(0)|$, i.e. $|x(t)-\mathbf{1}_{n}c(0)|\leq e^{-\alpha t}|\delta(0)|$. Hence, the switched multi-agent system (\ref{SMAS}) can achieve the exponentially consensus under arbitrary switching. $\blacksquare$

\begin{remark}\label{1}
In fact, there are some simple bounds that do not need to compute the Laplacian spectrum for sampling period $h$. For example, we have $\lambda_n\leq2\max_{i\in\mathcal{I}_{n}}\{d_{ii}\}$ by Ger$\breve{s}$gorin Disc theorem. Thus, the consensus problem of switched multi-agent system (\ref{SMAS}) can be solved if $\mathscr{G}$ is a undirected connected graph and the sampling period $0<h<\frac{1}{\max_{i\in\mathcal{I}_{n}}\{d_{ii}\}}$.
\end{remark}

Next, we consider the consensus of switched multi-agent system (\ref{SMAS}) in directed graph with fixed topology, i.e., $\mathscr{L}(t)=\mathscr{L}$ for any time $t$. A key lemma is given which is summary of the work in \cite{ren05,guo13}

\begin{lemma}\label{L-directed}
Consider a directed graph with fixed topology. Then, the CT multi-agent system (\ref{SMAS-CT}) can solve the consensus problem if and only if the directed graph $\mathscr{G}$ has a directed spanning tree, and the DT multi-agent system (\ref{SMAS-DT}) can solve the consensus problem if and only if the directed graph $\mathscr{G}$ has a directed spanning tree and $0<h<\min_{i=1,2,\cdots,n}\frac{2Re(\lambda_{i})}{|\lambda_{i}|^{2}}$. The consensus state is $\mathbf{1}_{n}w^{T}x(0)$, where $w^{T}\mathscr{L}=0$ and $w^{T}\mathbf{1}_{n}=1$.
\end{lemma}

\begin{theorem}\label{T-directed-fixed}
Suppose the communication network $\mathscr{G}$ is a directed graph and has a directed spanning tree. Then, the switched multi-agent system (\ref{SMAS}) can solve the consensus problem under arbitrary switching if the sampling period $0<h<\min_{i=1,2,\cdots,n}\frac{2Re(\lambda_{i})}{|\lambda_{i}|^{2}}$.
\end{theorem}
{\bf Proof.} Let $t=t_{c}+t_{d}$, where $t_{c}\in\mathbb{R}$ is the total duration time on CT subsystem (\ref{SMAS-CT}) and $t_{d}\in\mathbb{Z}$ is the total duration time on DT subsystem (\ref{SMAS-DT}). Due to $\mathscr{L}(t)=\mathscr{L}$ for any time $t$, we have
\begin{equation}\label{solution}
   \begin{aligned}
  x(t)=\exp(-\mathscr{L}t_{c})(I_{n}-h\mathscr{L})^{t_{d}}x(0).
   \end{aligned}
\end{equation}
Because the communication network $\mathscr{G}$ has a directed spanning tree, from Lemma \ref{L-directed}, we know that $\lim_{t_{c}\rightarrow\infty}\exp(-\mathscr{L}t_{c})=\mathbf{1}_{n}w^{T}$ and $\lim_{t_{d}\rightarrow\infty}(I_{n}-h\mathscr{L})^{t_{d}}=\mathbf{1}_{n}w^{T}$ if  $0<h<\min_{i=1,2,\cdots,n}\frac{2Re(\lambda_{i})}{|\lambda_{i}|^{2}}$, where $w^{T}\mathscr{L}=0$ and $w^{T}\mathbf{1}_{n}=1$.

When $t\rightarrow\infty$, there are three cases: \\
i) $t_{c}\rightarrow\infty$ and $t_{d}\in\mathbb{Z}$ is a constant. Thus, from (\ref{solution}), we have
\[
\begin{aligned}
 \lim_{t\rightarrow\infty}x(t)=\lim_{t_{c}\rightarrow\infty}\exp(-\mathscr{L}t_{c})(I_{n}-h\mathscr{L})^{t_{d}}x(0)
  =\mathbf{1}_{n}w^{T}(I_{n}-h\mathscr{L})^{t_{d}}x(0)=\mathbf{1}_{n}w^{T}x(0).
\end{aligned}
\]
ii) $t_{c}\in\mathbb{R}$ is a constant and $t_{d}\rightarrow\infty$. Thus,
\[
\begin{aligned}
 \lim_{t\rightarrow\infty}x(t)=\exp(-\mathscr{L}t_{c})\mathbf{1}_{n}w^{T}x(0)=\left(I_{n}+(-\mathscr{L}t_{c})+\frac{1}{2!}(-\mathscr{L}t_{c})^{2}+\cdots \right)\mathbf{1}_{n}w^{T}x(0)=\mathbf{1}_{n}w^{T}x(0).
\end{aligned}
\]
iii) $t_{c}\rightarrow\infty$ and $t_{d}\rightarrow\infty$.
\[
\begin{aligned}
 \lim_{t\rightarrow\infty}x(t)=\mathbf{1}_{n}w^{T}\mathbf{1}_{n}w^{T}x(0)=\mathbf{1}_{n}w^{T}x(0).
\end{aligned}
\]

Therefore, the switched multi-agent system (\ref{SMAS}) can solve the consensus problem under arbitrary switching. $\blacksquare$

\begin{remark}\label{2}
Note that the switched multi-agent system (\ref{SMAS}) presents a unified framework both the CT multi-agent system (\ref{m-CT}) and the DT multi-agent system (\ref{m-DT}). When the duration time of CT subsystem (\ref{SMAS-CT}) $t_{c}=0$, the distributed coordination of switched multi-agent system (\ref{SMAS}) becomes the distributed coordination of DT multi-agent system (\ref{m-DT}). And when $t_{d}=0$, the switched multi-agent system (\ref{SMAS}) becomes the CT multi-agent system (\ref{m-CT}).
\end{remark}

In the following, we consider the switched multi-agent system (\ref{SMAS}) in undirected graph with switching topologies $\{\mathscr{G}_{s}: s=\sigma(t)\in\mathscr{J}_{0}\}$, where $\mathscr{J}_{0}$ is a finite index set and $\sigma(t)$ is a switching signal that determines the network topology.

\begin{theorem}\label{T-undirected-switching}
Suppose the communication network $\mathscr{G}_{s}$ is undirected and connected for each $s\in\mathscr{J}_{0}$. Then, the switched multi-agent system (\ref{SMAS}) can solve the consensus problem if the sampling period $0<h<\min_{s\in\mathscr{J}_{0}}\frac{2}{\lambda_n(\mathscr{L}_{s})}$.
\end{theorem}
{\bf Proof.} Because the communication network $\mathscr{G}_{s}$ is undirected and connected for each $s\in\mathscr{J}_{0}$, similar to Theorem \ref{T-undirected-fixed}, we know that $c(t)=\frac{1}{n}\sum_{i=1}^{n}x_{i}(t)$ is time-invariant. Let $\delta(t)=x(t)-\mathbf{1}_{n}c(t)$ and $V(\delta(t))=\delta^{T}(t)\delta(t)$, we have
\[
\begin{aligned}
 \dot{ V}(\delta(t))=-2\delta(t)^{T}\mathscr{L}_{s}\delta(t)\leq-2\lambda_2(\mathscr{L}_{s})V(\delta(t))\leq -2k_{1}V(\delta(t))
\end{aligned}
\]
and
\[
\begin{aligned}
 &V(\delta(t+1))-V(\delta(t))=\delta^{T}(t)(I_{n}-h\mathscr{L}_{s})^{2}\delta(t)-\delta^{T}(t)\delta(t)\\
 &\leq (-2h\lambda_2(\mathscr{L}_{s})+h^{2}\lambda_2^{2}(\mathscr{L}_{s}))V(\delta(t))\leq k_{2}V(\delta(t)),
\end{aligned}
\]
where $k_{1}=\min_{s\in\mathscr{J}_{0}}\{\lambda_2(\mathscr{L}_{s})\}$,  $k_{2}=\min_{s\in\mathscr{J}_{0}}\{-2h\lambda_2(\mathscr{L}_{s})+h^{2}\lambda_2^{2}(\mathscr{L}_{s})\}$.
Due to $0<h<\min_{s\in\mathscr{J}_{0}}\frac{2}{\lambda_n(\mathscr{L}_{s})}$, we have $k_{1}>0$ and $-1<k_{2}<0$. For any $t=t_{c}+t_{d}$,
\[
\begin{aligned}
 V(\delta(t))&\leq e^{-2k_{1}t_{c}}(k_{2}+1)^{t_{d}}V(\delta(0))\leq e^{-2\beta t}V(\delta(0),
\end{aligned}
\]
where $\beta=\min\{k_{1}, \frac{ln(1/(k_{2}+1))}{2}\}$. Hence, the switched multi-agent system (\ref{SMAS}) can achieve the exponentially consensus. $\blacksquare$

We consider the following nonlinear consensus protocol as follows
\begin{equation}\label{nonlinear consensus protocol}
u_{i}(t)=
\left\{
   \begin{aligned}
   &\sum_{j=1}^{n}a_{ij}(t)f(x_{j}(t)-x_{i}(t)) &&for~~ CT~~ subsystem,\\
   &h\sum_{j=1}^{n}a_{ij}(t)f(x_{j}(t)-x_{i}(t)), &&for~~ DT~~ subsystem,  \\
   \end{aligned}
   \right.
  \end{equation}
where $\mathscr{A}=[a_{ij}(t)]_{n\times n}$ is the weighted adjacency matrix associated with the graph $\mathscr{G}(t)$ at time instant $t$, $h>0$ is the sampling period. Suppose that function $f:\mathbb{R}\rightarrow \mathbb{R}$ satisfies the following assumptions:\\
(1) $f(x)=0$ if and only if $x=0$;\\
(2) $f(x)$ is an odd function;\\
(3) $\gamma_{1}x\leq f(x)\leq\gamma_{2}x$, where $\gamma_{2}>\gamma_{1}>0$, for any $x\in\mathbb{R}^{+}$.

Similar to the proof of Theorem \ref{T-undirected-fixed} and Theorem \ref{T-undirected-switching}, we can obtain the following corollary.

\begin{corollary}\label{nonlinear control input}
Suppose the communication network $\mathscr{G}_{s}$ is undirected and connected for each $s\in\mathscr{J}_{0}$. Then, the switched multi-agent system (\ref{m-CT}--\ref{m-DT}) with nonlinear consensus protocol (\ref{nonlinear consensus protocol}) can solve the consensus problem if the sampling period $0<h<\frac{\gamma_{1}}{\gamma_{2}^{2}}\min_{s\in\mathscr{J}_{0}}\frac{2}{\lambda_n(\mathscr{L}_{s})}$.
 $\blacksquare$
\end{corollary}

\section{\bf Simulations }\label{s-Simulations}

In this section, we provided some simulations to demonstrate the effectiveness of the theoretical results in this paper.

\begin{example}\label{e-directed}

\begin{figure}
 \centering
  \includegraphics[width=5.00cm]{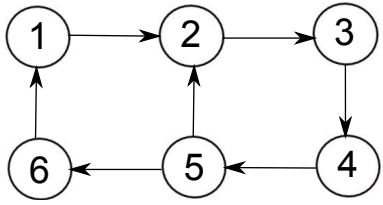}\\
  \caption{A directed graph $\mathscr{G}$ which has a directed spanning tree.}\label{fig1}
\end{figure}
When the communication network is chosen as in Fig. \ref{fig1}. It can be noted that $\mathscr{G}$ has a directed spanning tree. By calculation,
the  sampling period should satisfy $h<0.8112$. We choose $h=0.4056$. The switching law of switched multi-agent system (\ref{SMAS}) is shown in the top of Fig. \ref{fig2}. The state trajectories of all the agents are shown in the bottom of Fig. \ref{fig2}, which is consistent with the results in Theorem \ref{T-directed-fixed}.

\begin{figure}
 \centering
  \includegraphics[width=10.00cm]{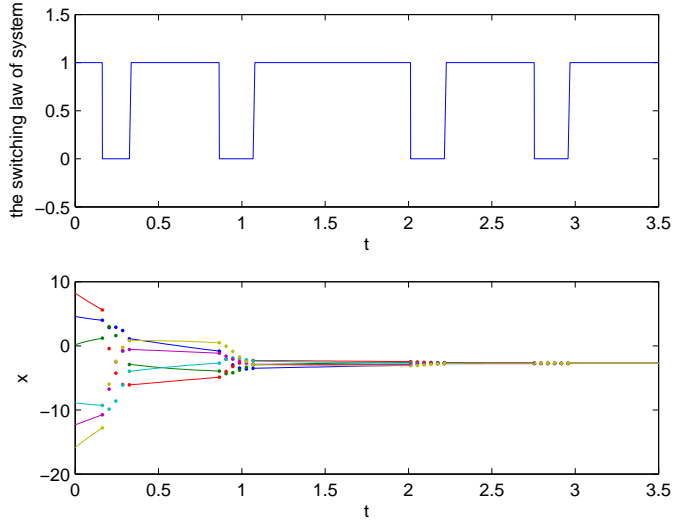}\\
  \caption{The switching law of system (\ref{SMAS}) and the state trajectories of all the agents.}\label{fig2}
\end{figure}

\end{example}

\begin{example}\label{e-switching}

\begin{figure}
 \centering
  \includegraphics[width=12.00cm]{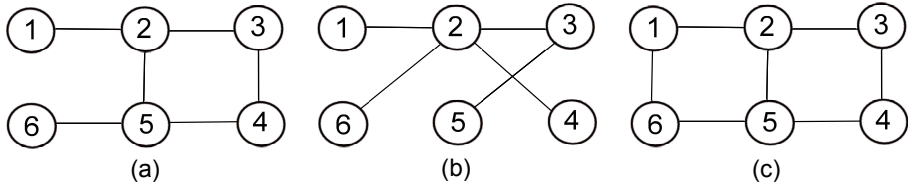}\\
  \caption{Three undirected connected graph.}\label{fig3}
\end{figure}
Suppose the communication network is chosen as in Fig. \ref{fig3}. Noted that $\mathscr{G}_{s}$ is undirected and connected for each $s\in\{a, b, c\}$. By calculation, the  sampling period should satisfy $h<0.4155$. We choose $h=0.01$. The switching law of network depicted in Fig. \ref{fig3} is shown in the top of Fig. \ref{fig4}. The switching law of switched multi-agent system (\ref{SMAS}) is shown in the middle of Fig. \ref{fig4}. The state trajectories of all the agents are shown in the bottom of Fig. \ref{fig2}, which is consistent with the results in Theorem \ref{T-undirected-switching}.

\begin{figure}
 \centering
  \includegraphics[width=10.00cm]{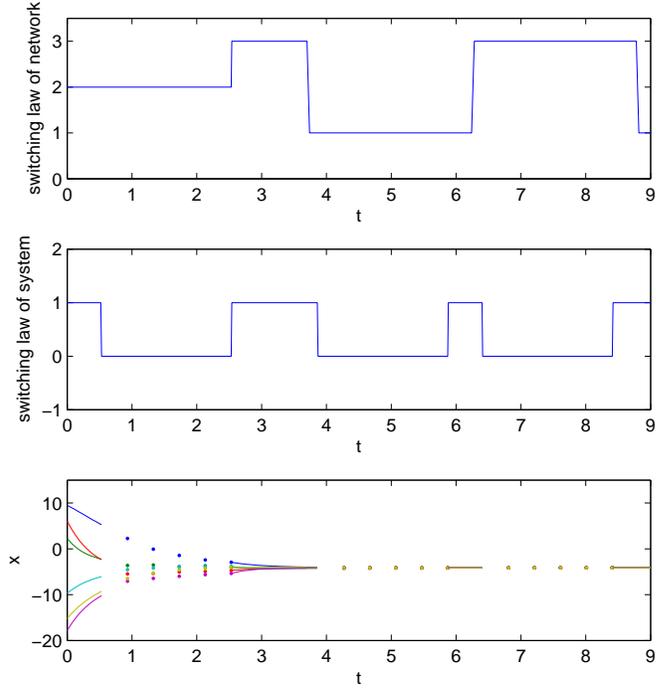}\\
  \caption{The switching law of network depicted in Fig. \ref{fig3}, the switching law of system (\ref{SMAS}) and the state trajectories of all the agents.}\label{fig4}
\end{figure}

\end{example}

\section{\bf Conclusions}\label{s-Conclusion}
In this paper, the consensus problem of switched multi-agent system which is composed of continuous-time and discrete-time subsystems is considered. The linear protocol is presented for solving the consensus problem. If the sampling period $0<h<\min_{i=1,2,\cdots,n}\frac{2Re(\lambda_{i})}{|\lambda_{i}|^{2}}$, we prove that the switched multi-agent system can achieve the consensus under undirected connected graph and directed graph, respectively. For switching topologies, a sufficient condition is also given if the communication network $\mathscr{G}_{s}$ is undirected and connected for each $s\in\mathscr{J}_{0}$. The future work will focus on the consensus of switched multi-agent system with time-delays, the containment control of switched multi-agent system etc.


\end{document}